\documentstyle[12pt]{article}
\begin{document}
\newcommand{\beq}{\begin{equation}}
\newcommand{\eeq}{\end{equation}}
\newcommand{\bea}{\begin{eqnarray}}
\newcommand{\eea}{\end{eqnarray}}
\textwidth=5truein
\textheight=7.8truein
\thispagestyle{empty}
\setcounter{page}{1}

\centerline{\bf LAGRANGIAN MODELS OF PARTICLES WITH SPIN:}
\vspace*{0.035truein}
\centerline{{\bf THE FIRST SEVENTY YEARS }\footnote{Dedicated to Jurek
Lukierski on the occasion of his 60$^{th}$ anniversary}}
\vspace{0.37truein}
\begin{center}
{\sc Andrzej Frydryszak}
\end{center}
\vspace*{0.015truein}
\centerline{\it Institute of Theoretical Physics, University of Wroc{\l}aw,}
\centerline{\it 50-204 Wroc{\l}aw, Pl. M. Borna 9, Poland}
\vspace{0.225truein}
\noindent {\bf Abstracts:}
We briefly review models of relativistic particles with spin.
Departuring from the oldest attempts to describe the spin within the
lagrangian framework we pass throught various non supersymmetric models.
Then the component and superfield formulations of the spinning particle
and superparticle models are reviewed. Our focus is mainly on the classical
side of the problem, but some quantization questions are mentioned as
well.\\
\vspace{2mm}
\baselineskip=15pt
\section{Introduction}
The aim of the present brief review is to indicate some essential
aspects of the theory of relativistic point particle with spin. The
selected
models are presented mostly historically as they were appearing, to show the
development of ideas in the period of 70 years - starting from the very
begining.
\par The first published work  concerning the lagrangian description of the
relativistic particle with spin was the paper by Frenkel which appeared
in 1926. In that time main considerations go towards the derivation
of the equations of spin precession in the external electromagnetic field.
Then to its relativistic generalization. The last one was achieved by
Bargmann, Michael and Telegdi \cite{bmt} 33 years after the Frenkel model was
constructed. However only the work of Frenkel contains the Lagrangian
defining the model of a particle, not only a considerations on the
equations of motion level. Then there is a forty years long gap in the
activity in constructing such a models. However, in the mean time - in
the fifties, the idea of anticommuting coordinates emerges in works of
Martin \cite{ma}, Matthews and Salam \cite{ms}, and Tobocman \cite{tob}.
Later it strongly influenced particle models \cite{{1001},{sund}}.\\
The silence was broken with the work of Barut \cite{bar}. Then in early seventies
wider interest in the subject begins with the works of Hanson,
Regge \cite{hr}; Grassberger \cite{gra}; Casalbuoni \cite{cas};
Berezin, Marinov \cite{bm}. Some of these models involve
anticommuting coordinates. Further growth of the interest was stimulated
by dynamically developing research in the supersymmetry and supergravity
and then superstrings theory with the wide use of {\bf Z}$_2$-graded
structures. This period lasts from the eighties to the present decade with
such a new models: Brink-Schwarz \cite{bs}, Brink-diVecchia-Howe
\cite{bdh}, de Azc\'{a}rraga-Lukierski \cite{al1}, Siegel
\cite{{sieg1},{sieg2}}, Volkov-Soroka-Tkach \cite{stv1}.
Above metioned models fall in principe into different categories. The
classification can be made due to the such attributes as mass,
algebraical (conventional or anti-commuting) and geometrical character of
the internal degrees of freeedom (vectorial, spinorial, twistorial).
\par In the sequel we shall  adopt the
following naming conventions. Models involving only conventional
coordinates will be called the classical models. They in principle are
of vectorial or tensorial type. The models involving anti-commuting
coordinates are generally called here pseudoclassical. These with the
anti-\-commuting vectorial degrees of freedom are called the
spinning particles and these with the spinorial anti-commuting degrees of
freedom are called superparticles.

\par The type of extension of the configuration space of the relativistic
particle by the commuting or anti-commuting coordinates determines the
symmetry and the behaviour of the model in the external field and upon
quantization.
Starting demand is to have object which is at least Poincar\'{e}
invariant. Classical vectorial particles and spinnig particles couple
properly to the external fields, however only the latter ones can be
correctly quantized and do not give undesirable classical
selfacceleration and effect of
Zitterbewegung type. Spining particles are in some sense the classical
limit of the Dirac particle.
After the first quantization these new anti-commuting
variables are mapped into the Dirac matrices and they disappear from the
theory. This is a general feature of the spinning particle models.
\par On the other hand the extension of the configuration space by the
anti-commuting spinorial variables yields the models which are
super-Poincar\'{e} invariant. In contrast to the spinning particle
models their first quantization gives theory which still involves the
anti-commuting variables. As a result of quantization we get rather
not a single quantum particle with spin $\frac{1}{2}$ but a minimal
supermultiplet.
\par The organization of the review follows the models classification
sketched above. We begin with the two principal categories of the
classical models and the so called arbitrary spin particles.
Then the pseudoclassical group of models is presented including spinning
particles, super-particles, twistorial and harmonic particles, arbitrary
superspin models.
Next we comment the double supersymmetric models with the
spinning superparticle in the component and superfield form.
We conclude this brief review recapitulating some new developments
including first attempts of q-deformation the relativistic model of the
spinnig particle and the $\kappa$-relativistic model of a particle.
\par The literature on the particle with spin is vast. We include here,
only the very selective list of references.
\section{Classical models}
In this section I shall briefly present the classical models of the
relativistic particle with spin. The adjective classical means here not
only that a model is not quantum but also that it is described by means
of the commuting variables only.
\subsection{Vectorial models}
\par Historically the first model has been introduced by Frenkel. The
spin of the particle in his model is described directly by a
tensor of spin $S_{\mu \nu}$,
which is assumed to be proportional to the tensor of internal
magnetic moments $M_{\mu \nu }$. It enters the lagrangian via the
"transversality condition" 
\beq
S_{\mu \nu }\dot{x}^{\nu} = 0.
\eeq
to reduce the number of independent degrees of freedom.
We shall call it the Frenkel condition. Explicit form of the action is
as follows
\beq
S=\int d\tau (\lambda \dot{x}^2 + a^{\mu}S_{\mu \nu}\dot{x}^{\nu} + S_{\mu \nu}
{\omega}_{\mu \nu}), \quad {\omega}_{\mu \nu} = -{\omega}_{\nu \mu}
\eeq
It yields the equations of the motion of the form
\bea
\dot{S}_{\mu \nu}-(\dot{x}_{\mu}S_{\nu \rho} - \dot{x}_{\nu}
S_{\mu \rho})a^{\rho}&=&0\\
(\lambda x_{\mu} + S_{\mu \rho}a^{\rho})^{.} &=& 0
\eea
Some developments of this model were done 33 years later by Barut. To
describe internal degrees of freedom he introduces the frame of four
fourvectors $q^{\mu}_{(i)}$ $i=0,1,2,3$; such that $q^{\mu}_{(0)}$ is
proportional to $(\dot{x}^{\mu})$ and the rest is orthogonal to
$(\dot{x}^{\mu})$. Using implicit form of the action
\beq
S=\int d\tau L(\dot{x}^{\mu}, q^{\mu}_{(i)}, \dot{q}^{\mu}_{(i)})
\eeq
and the following definition of the tensor of spin
\beq
S_{\mu \nu}=\frac{\partial L}{\partial \dot{q}^{\mu}_{(i)}}q^{\nu}_{(i)} -
\frac{\partial L}{\partial \dot{q}^{\nu}_{(i)}}q^{\mu}_{(i)}
\eeq
he gets the following form of the equations of motion
\bea
\dot{p}^{\mu}&=&0\\
\dot{S}_{\mu \nu}+p_{[ \mu} x_{\nu ]} &=& 0
\eea
Moreover the Frenkel condition is valid.\\
Historically the next classical model was of different kind. It was an
exemplification of the idea that particle is an irreducible
representation of the Poincar\'{e} group. Namely in the model of Hanson
and Regge the configuration space has coordinates $(x_{\mu},
\Lambda_{\mu \nu})$, where $\Lambda \in L^{\uparrow}$ ($L^{\uparrow}$ -
orthochronous Lorentz group). It turns out that dependence of the
lagrangian function on the $\dot{x}$, $\Lambda$, $\dot{\Lambda}$ should
be restricted to $L(\dot{x}^{\mu}, {\sigma}^{\mu \nu})$, where
${\sigma}^{\mu \nu}={\Lambda}^{\lambda \mu}\dot{\Lambda}_{\lambda}^{\nu}$.
Now the tensor of spin is given by the formula
\beq
S^{\mu \nu} = \frac{\partial L}{\partial {\sigma}^{\mu \nu}}
\eeq
and the equations of motion take the same form as in the Barut model.
The demand, that in the non-\-relativistic limit the particle has only
three spin degrees of freedom is realized by the condition
\beq
S_{\mu \nu}p^{\nu} = 0
\eeq
It was originally introduced by Dixon \cite{dix}. In this model it should be
included into the action. Let us note that the Frenkel condition and
Dixon condition yield the essential differences in possible motions of
particles, even in the free case. Explicit realization of such an action
takes the form
\bea
L(\dot{x}^{\mu}, {\sigma}^{\mu \nu})&=& -\frac{1}{\sqrt{2}}\{
A\dot{x}^2-B\sigma^2+[(A\dot{x}^2-B\sigma^2 )^2 \\
&-& 8B(A\dot{x}\sigma^2\dot{x} - 2Bdet\sigma )]^{\frac{1}{2}}
\}^{\frac{1}{2}},
\eea
where ${\sigma}^2={\sigma}_{\mu \nu}{\sigma}^{\mu \nu}$ and
$\dot{x}{\sigma}^2\dot{x}=\dot{x}_{\mu}{\sigma}^{\mu \nu}
{\sigma}_{\nu \lambda}\dot{x}^{\lambda}$. The constraints are of the
form
\bea
S_{\mu \nu}p^{\nu} &=& 0\\
p^2 -\frac{A}{4}S_{\mu \nu}S^{\mu \nu}&=& 0
\eea
The last constraint follows from the reparametrization invariance of the
action. The mass of the particle is renormalized here by the square of
the $S^{\mu \nu}$. This model does not give after the first quantization
the Dirac particle.\\
Along similar lines is constructed the BMSS model proposed in Ref. \cite{mbss}.
Here again the particle with spin is directly tied up to the irreducible
representations of the Poincar\'{e} group $P^{\uparrow}_{+}$. To this
end, as a configuration space one takes $(z^{\mu},\Lambda )\in
P^{\uparrow}_{+}$. The matrix is decomposed into the momentum and spin
tensor, where
\bea
p_{\mu} &=& m{\Lambda}_{\mu 0},\quad m 0 \\
S &=& i\lambda \Lambda{\sigma}_{12}{\Lambda}^{-1}, \lambda \in \Re \\
({\sigma}^{\mu \nu})_{\rho \lambda} &=& -i {\delta}^{\mu \nu}_{\rho \lambda}
\eea
This means that
\beq
S_{\mu \nu} = \lambda({\Lambda}_{\mu 1}{\Lambda}_{\nu 2} -
{\Lambda}_{\mu 2}{\Lambda}_{\nu 1})
\eeq
and by the construction
\bea
S_{\mu \nu}S^{\mu \nu}&=& 2{\lambda}^2\\
p^{\mu}S^{\mu \nu}&=& 0
\eea
Therefore the model written on such a space has to be of Dixon category.
The lagrangian finally defining the model is taken in the following form
\beq
L = p_{\mu}\dot{z}^{\mu} + \frac{i\lambda}{2}
Tr({\sigma}_{12}{\Lambda}^{-1}\dot{\Lambda})
\eeq
The resulting equations of motion are the same as in the Hanson-\-Regge
model. The four-\-momenta and four-\-velocities are related in the
standard way. This type of model has been recently reformulated and a
correspondence to the psudoclassical model was proposed \cite{ho1}.\\
Now let us come back to the vectorial models. In 1978 Grassberger
proposed the description \cite{gra} in which the Minkowski space is extended by
the two four-\-vectorial internal degrees of freedom $(x_{\mu})\mapsto
(x_{\mu}, a_{\mu}, b_{\mu})$. The Poincar\'{e} invariant action is
defined by means of the lagrangian
\beq
L= \frac{1}{2}m(1-{\dot{x}}^2 )+\dot{x}_{\mu}(\beta b^{\mu}-\alpha
a^{\mu})+\frac{1}{2}({\dot{b}}_{\mu}a^{\mu}-{\dot{a}}_{\mu}b^{\mu})
\eeq
The Lagrange multipliers $m$, $\alpha$, $\beta$ are introduced to provide
the necessary constraints
\bea
{\dot{x}}^2 &=& 1\\
a_{\mu}{\dot{x}}^{\mu}&=& 0 \\
b_{\mu}{\dot{x}}^{\mu}&=& 0
\eea
The tensor of spin obtained from the above lagrangian is composed of the new
vectorial internal co-\-ordinates i.e.
\beq
S_{\mu \nu}= a_{\mu} b_{\nu} - a_{\nu} b_{\mu}
\eeq
and it obeys the Frenkel condition with $S_{\mu \nu}S^{\mu \nu}=const$.
Now there are some new features in the equations of motion of this model,
namely
\bea
\frac{d}{d\tau}(m{\dot{x}}^{\mu}+{\eta}^{\mu})&=& 0\\
{\dot{S}}_{\mu \nu}+p_{[ \nu}{\dot{x}}_{\nu ]}&=& 0, \quad
{\eta}_{\nu}=\alpha a_{\nu}+\beta b_{\nu}
\eea
This means that if such a particle is coupled to the external
electromagnetic field, or has only passed through the bounded area with
non-vanishing field, due to the presence of ${\eta}_{\nu}$ term the
center of mass and the center of the charge need not coincide.\\
Five years later Cognola, Soldati, Vanzo and Zerbini \cite{csvz} proposed another
vectorial model. Its configuration space is the same as for the
Grassberger model, but the new lagrangian takes the form
\beq
L= -m^2(P_{\mu \nu}{\dot{x}}^{\mu}{\dot{x}}^{\nu})^{\frac{1}{2}}
-{\dot{a}}_{\nu}b^{\nu}
\eeq
with $S_{\mu \nu}$ given by eq.(26) and
\beq
P_{\mu \nu}=g_{\mu \nu}-\frac{S_{\mu \rho}S_{\nu}^{\rho}}{S^2} =
g_{\mu \nu}-Q_{\mu \nu}
\eeq
The constraints are now of the form
\bea
p^2&=&m^2, \quad p_{\mu}=\frac{m}{\sqrt{P_{\mu \nu}{\dot{x}}^{\mu}
{\dot{x}}^{\nu}}}({\dot{x}}_{\mu}-Q_{\mu \nu}{\dot{x}}^{\nu})\\
p_{\mu}a^{\mu}&=&0\\
p_{\mu}b^{\mu}&=&0
\eea
what obviously means that the Dixon condition is fulfilled. The internal
degrees of freedom are here othogonal to the momenta, moreover the
direction of fourvelocity and fourmomenta can be different.
Equations of motion take the standard form.\\
Another branch of the models of the classical spinnig particle is
connected with the Souriau's notion of the space of motions \cite{sou}
and the coadjoint orbit method. As a sample we indicate here the Refs.
\cite{{sou},{duvhor},{duv1}} and recently \cite{zakrz}.\\
The classical models sketched in this section have one property in
common, they do not give after the first quantization the accepted
quantum relativistic Dirac particle. On the other side, coupled to the
external electromagnetic field, in the limit of the weak homogenous
field, they yield the Bargmann-Michel-Telegdi equations. The models
fulfilling the Frenkel condition have helisoidal curves as the solutions
of the equations of motion. This can be interpreted as a counterpart of
the Zitterbewegung solution for the  Dirac particle, however from the
classical point of view such a trajectory for a free particle can be
hardly accepted.\\
Despite the technical subtleties of different models of this kind the
behaviour of the particular type of the particle with spin depends
mainly on the type of the "othogonality condition" for the internal
degrees of freedom i.e. the Frenkel or the Dixon condition. The latter
one seems to be more natural.
\subsection{Spinorial models (arbitrary spin particles)}
Finally let us comment classical models of the particle with spin
described by the spinorial coordinates. The presence of the commuting
spinor not necessarily means that model describes the particle with spin
\cite{{pmc},{bbcl}}. In the twistor-\-like approach the
massless point particle for example, has the action of the form
\beq
S=\int d\tau p_m ({\dot{x}}^m -\bar{\lambda} {\gamma}^m \lambda ),
\eeq
where $p_m$ is the particle momentum and ${\lambda}^{\alpha}$ is a
commuting spinorial variable, needed to ensure the mass shell condition
(another interesting spinorial model  has been discussed in Refs.\cite{fer}
and \cite{shi} with the action of the form $S=\int
d\tau\bar{\lambda}{\gamma}_m \lambda {\dot{x}}^m )$.\\
The bove action is a good starting point to supersymmetric generalizations.
There are models of point particles with spin described by commuting
spinors, however in such an approach not only the spin one half appears but
also the whole spectrum of spin values. This justifies the name:
arbitrary spin particles \cite{{peri},{getk},{hppt},{has},{kuz}}.
As an ilustration let us consider two models of the arbitrary spin
massive particles. The lagrangian of the first model \cite{has} is
closely related to the spinning superparicle model \cite{aklh}, and is
given in the form
\beq
L=\frac{1}{2} (e^{-1} {\dot{x}}^2 + e m^2_0 ) - h \dot{x}\cdot j +
2\bar{\eta} \dot{\eta},
\eeq
where $h, m_0 \in {\Re}_+$ and $\eta$ is the Majorana spinor. The
current $j^a =\bar{\eta}{\gamma}^a \eta$ has vanishing square. The
resulting equation of motion have the form
\bea
\frac{d}{dt} ( e^{-1} \dot{x} - hj ) &=& 0 \\
{\dot{x}}^2 + m_0^2 c^2 &=& 0 \\
h \dot{x}\eta - 2\dot{\eta} &=& 0
\eea
Obviously the conserved angular momentum tensor has a contribution from
the "internal" degrees of freedom
\beq
M_{ab}=p_b x_a -p_a x_b -\bar{\eta}{\gamma}_{ab}\eta ,
\eeq
where $p_a$ is defined by expression in the first equation of motion
given above. The Dirac quantization of this particle, after solving
the second class constraints (and hence with the breaking of the Lorentz
covariance in the spinorial sector of the phase space) gives the
condition on states which singles out arbitrary spin and relates the
mass and the spin
\beq
m_{J\pm } = \pm h \frac{J+1}{2} + \sqrt{h^2
(J+\frac{1}{2})^2 + m_0 }
\eeq
The limit for the massless case can be considered as well and gives the
description of particles with arbitrary helicity.\\
The second example of the arbitrary spin particle model comes \cite{kuz}
from the geometrical construction of the model on the six dimensional
product space of the Minkowski space $M$ and the two-\-dimensional
sphere $S^2$. The family of Lagrangians of the model involves the joint
interval in $M$ and $S^2$, with the metric on $S^2$ depending explicitly
on fourvelocities. It is parametrized by the mass and spin $(m,s)$ and
has the following form
\beq
L=\frac{1}{2} e^{-1}_1 ({\dot{x}}^2 - (e_1 m c)^2 )+e^{-1}_2
(4\frac{z\dot{\bar{x}}}{(\dot{x}\cdot \xi )^2}e^2_1 + (\Delta e_2 )^2 ),
\eeq
where $z$ is a complex coordinate on the $S^2$, $e_1$, $e_2$ are einbein
fields associated with the reparametrizations in $M$ and $S^2$. The
$\Delta$ is an additional ("spherical") mass $\Delta = \hbar mc
\sqrt{s(s+1)}$ (s - spin) . The conserved momentum tensor has the form
\beq
M_{ab} = x_a p_b -x_b p_a
+({\bar{z}}^{\dot{\alpha}}({\tilde{\sigma}}_{ab})_{\dot{\alpha}\dot{\beta}}
{\bar{z}}^{\dot{\beta}} p_{\bar{z}} - z^{\alpha}({\sigma}_{ab})_{\alpha \beta}
z^{\beta} p_z )
\eeq
with the "internal" part defined by complex spherical coordinates. The
Dirac quantization of this model gives, like in the previous one, the
whole spectrum of spins. Let us note that this model exhibits the
Zitterbevegung effect which is typical for the vectorial
models.\\  
\section{Pseudoclassical models}
In this section we pass to the models with the internal degrees of
freedom described by the anti-\-commuting co-\-ordinates. The origins of
the pseudomechanics should be dated back to the 1956, to the work
of Martin \cite{ma}. However, anti-\-commuting variables appear firstly in the
context of the functional integral for fermions in the works of
Matthews, Salam \cite{ms} and Tobocman \cite{tob}.
The extension of a configuration space to
superspace enlarges the underlying symmetry group. Depending on the type
of the model the extension of the Poincar\'{e} algebra yields the
super-Poincar\'{e} algebra or some super algebra of the other kind. There
are two types of such a models: vectorial and spinorial.\\
In models of vectorial type (the spinning particle models) the extension
gives a untypical vectorial superalgebra, with the odd generators having
vectorial index. Characteristic feature of such models is conventional
character of the first quantized theory. Namely, the odd variables upon
quantization are mapped into Dirac matrices and disappear on the quantum
level. Such particle can be considered as a pseudoclassical limit of
the conventional Dirac quantum particle. In fact, this was the
non-achieved goal of the vectorial classical models presented in the
previous section.\\
The spinorial models (the superparticle models) have the super-Poincar\'{e}
algebra (or its extension) as a symmetry generators. This models are
connected more closely to the relativistic supersymmetry and the
superparticles can be viewed upon, as a minimal, irreducible representations
of the super-Poincar\'{e} group. However, such an objects contains the
whole multiplet of fields with different spin but not only the spin one
half component. On the first quantized level one still deals with the
anti-\-commuting variables and instead of the wave functions the
the wave super-functions have to be considered. The Dirac equation is
not used literary but finds its superspace counterpart. It is worth
noting that there exists an equivalence betveen some pseudoclassical and
classical models of particles with spin which allows to generalize the
notion of Zitterbevegung to the pseudoclassical case \cite{bap} (cf. as
well Ref. \cite{ho2}).
\subsection{Spinning particles}
Twenty years after the anti-\-commuting variables were introduced into
the physical literature for the first time , there was proposed the
spinning particle model by Berezin, Marinov \cite{bm} and Barducci, Casalbuoni,
Lusanna \cite{bcl}.\\
The configuration space for this model is described by the set of
co-\-ordinates $(x_{\mu}, {\theta}_{\mu}, {\theta}_5 )$, where the
$\theta$-variables are anticommuting between themselves;
${\theta}_{\mu}$ beeing fourvector and ${\theta}_5$ a scalar. Proposed
lagrangians were of the form
\bea
L_{BCL}&=&-m\sqrt{({\dot{x}}^{\mu}-\frac{i}{m}{\theta}^{\mu}{\dot{\theta}}_5)
({\dot{x}}_{\mu}-\frac{i}{m}{\theta}_{\mu}{\dot{\theta}}_5)}-\frac{i}{2}
{\theta}_{\mu}{\dot{\theta}}^{\mu}-\frac{i}{2}{\theta}_5{\dot{\theta}}_5\\
L_{BM}&=&-m\sqrt{{-\dot{x}}^2}+\frac{i}{2}\left({\theta}_{\mu}{\dot{\theta}}^{\mu}
+{\theta}_5{\dot{\theta}}_5-(\frac{{\dot{x}}^{\mu}}{\sqrt{{-\dot{x}}^2}}
{\theta}_{\mu}+{\theta}_5)\lambda\right)
\eea
Let us focus on the model given by the first of  above lagrangians. It
is invariant under the supertranslations
\bea
x_{\mu}\mapsto {x'}_{\mu}&=&x_{\mu}-{\epsilon}_{\mu}A{\theta}_5+{\epsilon}_5
B{\theta}_{\mu}\\
{\theta}_{\mu}\mapsto {{\theta}'}_{\mu}&=&{\theta}_{\mu}+{\epsilon}_{\mu}\\
{\theta}_{5}\mapsto {{\theta}'}_{5}&=&{\theta}_{5}+{\epsilon}_{5}
\eea
where $A$, $B$ are numerical constants. The algebra of generators of
these transformations is defined by the following relations
\bea
\{ Q_{\mu}, Q_{\nu} \} &=& ag_{\mu \nu}\\
\{ Q_{5}, Q_{5} \} &=& b\\
\{ Q_{\mu}, Q_{5} \} &=& (B-C)P_{\mu}
\eea
The $Q_{\mu}$, $Q_{5}$ commutes with $P_{\mu}$. Performing canonical
analysis of the model one gets the first class constraints
\bea
p^2-m^2&=&0\\
p_{\mu}{\theta}^{\mu}-m{\theta}_5&=&0
\eea
After the first quantization one obtains precisely the Klein-\-Gordon and
Dirac equations. For the $\theta$-sector of the phase space the
anticommutation relations
\bea
\left[ {\hat{\theta}}_{\mu}, {\hat{\theta}}_{\nu} \right]_+ &=& -h g_{\mu \nu}\\
\left[ {\hat{\theta}}_{5}, {\hat{\theta}}_{5} \right]_+ &=& \hbar\\
\left[ {\hat{\theta}}_{\mu}, {\hat{\theta}}_{5} \right]_+ &=& 0
\eea
show that the classical variables originating from the Grassmann algebra
are mapped after quantization to the elements of the appropriate
Clifford algebra, here ${\theta}_{\mu}\mapsto
\sqrt{\frac{\hbar}{2}}{\gamma}_{\mu}{\gamma}_{5}$,
${\theta}_{5}\mapsto \sqrt{\frac{\hbar}{2}}{\gamma}_{5}$
(where ${\gamma}_{\mu}$, ${\gamma}_{5}$ - Dirac matrices).\\
Above model can be generalized taking into account the reparametrization
invariance, which yields the supergravity in d=1 \cite{bcl} (cf. also J. van
Holten's contribution to this volume). For the relativistic point
particle one can explicitly achieve time reparametrization invariance by
means of einbein field $e(\tau )$. In the case of the spinning particle
it is necessary to introduce its supersymmetric partner $\psi(\tau )$.
Resulting lagrangian takes the form
\beq
L=e^{-1}{\dot{x}}^2+em^2+i({\theta}_{\mu}{\dot{\theta}}^{\mu}+
{\theta}_5{\dot{\theta}}_5 )+i(m{\theta}_5
-e^{-1}{\dot{x}}^{\mu}{\dot{\theta}}_{\mu})\psi
\eeq
The action given by this lagrangian is invariant under Poincar\'{e}
transformations, reparametrizations and the local supersymmetry
transformations, what justifies the associacion with the $D=1$
supergravity. Namely,
\bea
\delta\tau &=& -\alpha (\tau )\\
\delta x^{\mu} &=& \alpha (\tau ){\dot{x}}^{\mu}+i\epsilon (\tau
){\theta}^{\mu}\\
\delta {\theta}^{\mu} &=& \alpha (\tau ){\dot{\theta}}^{\mu}+(2e)^{-1}
\epsilon (\tau )(2{\dot{x}}^{\mu}-i\psi {\theta}^{\mu})\\
\delta {\theta}_{5} &=& \alpha (\tau ){\dot{\theta}}_{5}-m\epsilon (\tau
)\\
\delta e &=& \frac{d}{d\tau}(\alpha (\tau )e+i\epsilon (\tau ) \psi )\\
\delta \psi &=& \frac{d}{d\tau}(\alpha (\tau )\psi +2\epsilon (\tau ))
\eea
The Euler-\-Lagrange equations for the $e$ and $\psi$ are of algebraic
character and this fields can be easily eliminated what yields the other
version of the lagrangian which was considered in \cite{{gt},{sund}}. The
most general form of the action for the spinning particle with the
supergravity multiplet can be given by the lagrangian of the form
\cite{bggs}
\bea
L&=& ig e^{-1}{\dot{x}}^2 -ibe m^2+g{\theta}_{\mu}{\dot{\theta}}^{\mu}+
b{\theta}_5{\dot{\theta}}_5 +mb{\theta}_5\psi \\
&-&(ge^{-1}+2e^{-3}{\dot{x}}^2 g')\psi ({\dot{x}}^{\mu}{\dot{\theta}}_{\mu})
+2e^{-2}g'({\dot{x}}^{\mu}{\dot{\theta}}_{\mu})^2
\eea
All the spinning particle models have the property that they are the
classical limits of the Dirac field theory and the anticommuting
variables are present only in the classical description. The coupling of
such models to the external electromagnetic or Yang-\-Mills fields
yields vectorial superspace versions of the Bargmann-\-Michel-\-Telegdi
or Wong equations \cite{bcl}. Therefore in some sense the spinning particle
models are improved versions of the conventional vectorial models, now
with the proper quantum picture.\\

Let us finish this section with the superfield formulation of the
spinning particle proposed by Ikemori \cite{{ike1},{ike2}}. The first step
consists in considering instead of single conventional time parameter a
generalized super-\-time as a $(1|1)$ dimensional superspace with
coordinates $(t,\eta )$, where $\eta$ is the new anticommuting variable.
This means that trajectories of a system will take values in the
superspace too. Namely,
\beq
X(t,\eta) = x(t)+i\eta \theta (t), \quad \quad X\in C^{\infty}(t)[ \eta ]_0
\eeq
Above superfield unifies in one object the even and odd coordinates of
the spinning particle. The supersymmetry present in the super-time
space is called the little SUSY: $(t,\eta)\longrightarrow (t+\tau ,
\alpha \eta , \eta +\alpha )$, where $\tau$, is an even and $\alpha$ an
odd infinitesimal parameter. The algebra of supercharges and covariant
derivatives is of the form
\bea
&&Q= {\partial}_{\eta}+\eta {\partial}_t, \quad
{\partial}_{\eta}=\frac{\partial}{\partial \eta}, \quad
{\partial}_t=\frac{\partial}{\partial t}\\
&&D= {\partial}_{\eta}-\eta {\partial}_t ,\\ 
&&\left[ Q,Q \right]_{+} = 2 {\partial}_t \\
&&\left[ D,D \right]_{+} = -2 {\partial}_t \\
&&\left[ Q,D \right]_{+} = 0
\eea
To introduce the local invariance the $d=1$ supergravity multiplet is
needed. It enters the super-zweibein field $(E_A^M)$, where ${\partial}_M
=({\partial}_t , {\partial}_{\eta})$ and ${\nabla}_A = E_A^M
{\partial}_M$. The action takes the form
\beq
S= \frac{1}{2} \int dt d\eta sdet(E_A^M) g^{AB} E_A^M {\partial}_{M}
X^{\mu}E_B^N {\partial}_{N}X^{\nu}g_{\mu \nu}.
\eeq
The customary choice of
the gauge for the super-zweibein field is the following
\beq
(E_A^M)=\left(
\begin{array}{ll}
E^{-1}&-e^{-1}\psi \\
-e^{-1}\eta & e^{-1}E
\end{array}
\right), \quad \quad E=e+\eta \psi
\eeq

Recently the model of the spinning particle with arbitrary number of
supersymmetries on the world-\-line has been constructed
\cite{{gatn1},{gatn2}}.
Such an $N$-extended  little SUSY in the massive model of the
spinning particle, after the field redefinitions in the equations of
motion, yields the supersymmetric Lax equation. Moreover it can be used
in the study of hyperbolic Kac-\-Moody algebras.
\subsection{Superparticle models}
The extension of the Minkowski space to the superspace with the
additional spinorial coordinate is the basic structure of the
supersymmetric field theories \cite{1001}.\\
The super-Poincar\'{e} group becomes the fundamental symmetry of the
theory. Now, the superparticles are generalizations of the relativistic
point particle from the Minkowski space to such a superspace, with still
the same requirement at the background - to describe "correctly" the
spin. Because they incorporate super-Poincar\'{e} invariance there is a
close connection between superparticles and representation of
supersymmetry. Some models provide a natural examples of actions which
yield, after the first quantization, the minimal irreducible representations
of the given super-Poincar\'{e} superalgebra. \\
The superparticles have reach symmetry, as local as rigid \cite{tow}. In many
respects the quantization procedure is difficult because of the
complicated structure of the phase space constraints. This aspect of the
superparticle models makes that they are instructive toy models used to
understand the superstrings and the variety of their quantization
procedures.\\
There are important differences between massless and massive models,
however we will not stress them, aiming only to ilustrate generally the
historical development in the construction of the models.\\
The first pseudoclassical relativistic particle model with the spinorial
grassmanian co-\-ordinates was proposed by Casalbuoni in 1976
\cite{cas}. On the configuration superspace $(x_{\mu}$,
${\theta}_{\alpha}$, ${\bar{\theta}}^{\dot{\alpha}})$ the he defined the
lagrangian of the form
\beq
L=-m\sqrt{{\dot{\omega}}_{\mu}{\dot{\omega}}^{\mu}}
\eeq
where
\beq
d{\omega}_{\mu}=dx_{\mu}-i(d\theta {\sigma}_{\mu} \bar{\theta}
-\theta {\sigma}_{\mu} d\bar{\theta})
\eeq
is the super one-form, invariant under supertranslations
\bea
x_{\mu}\mapsto {x'}_{\mu}&=&x_{\mu}-i({\epsilon}{\sigma}_{\mu} \bar{\theta}-
\theta{\sigma}_{\mu}\bar{\epsilon})\\
{\theta}_{\alpha}\mapsto
{{\theta}'}_{\alpha}&=&{\theta}_{\alpha}+{\epsilon}_{\alpha}\\
{\bar{\theta}}_{\dot{\alpha}}\mapsto {{\bar{\theta}}'}_{\dot{\alpha}}&=&
{\bar{\theta}}_{\dot{\alpha}}+{\bar{\epsilon}}_{\dot{\alpha}}
\eea
The lagrangian is too poor to give after the first quantization the
Dirac equations and some interesting supersymmetric multiplets. The
chiral supermultiplet content of the Casalbuoni's G$_4$ model was
analysed by Almond \cite{alm}. To improve this model
the first order fermionic kinetic terms are needed.
But they cannot be introduced in a strightforward way, because of the
relation
\beq
{\theta}_{\alpha}\frac{d}{d\tau}{\theta}^{\alpha}=
\frac{d}{d\tau}({\theta}_{\alpha}{\theta}^{\alpha})
\eeq
Four years later Volkov and Pashnev \cite{vp} tried to cure this
drawback using more general super one-form, invariant under super-Poincar\'{e}
transformations. Namely,
\beq
ds^2=d{\omega}_{\mu}d{\omega}^{\mu}+ad{\theta}^{\alpha}d{\theta}_{\alpha}-
a^{\star}d{\bar{\theta}}^{\dot{\alpha}}d{\bar{\theta}}_{\dot{\alpha}},
\quad a\in C
\eeq
and then the action of the form
\beq
S=-m{\int}_{{\tau}_1}^{{\tau}_2}\sqrt{ds^2}=
-m{\int}_{{\tau}_1}^{{\tau}_2}\sqrt{{\dot{\omega}}_{\mu}{\dot{\omega}}^{\mu}
+a{\dot{\theta}}^{\alpha}{\dot{\theta}}_{\alpha}-
a^{\star}{\dot{\bar{\theta}}}^{\dot{\alpha}}
{\dot{\bar{\theta}}}_{\dot{\alpha}}}\tau
\eeq
Now the fermionic kinetic term is present and gives the first class
constraints. However, not the one playing upon quantization the
r\^{o}le of
the Dirac equation. Nevertheless the content of the model is more reach,
since the first quantized theory contains some multiplets (two scalar
multiplets and one vector multiplet of states with the negative norm).\\
The Brink-\-Schwarz action for a superparticle of mass $m$ in $d$
dimensions uses again invariant super oneform $\omega$.The
reparametrization invariance is provided by the einbein field what
enables to consider a massless superparticle as well. In 1981 they
proposed an action of the form \cite{bs}
\beq
S=\int d\tau (e^{-1}{\omega}^2 -em)
\eeq
where
\beq
{\omega}^n = {\dot{x}}^n +i\dot{\theta}{\Gamma}^n \theta ,\quad
n=1,2,...,d-1
\eeq
Specialy massless case is intersting here, because there is an
additional invariance present (Siegel \cite{sieg})
\bea
{\delta}_{\kappa}{\theta}^{\alpha}&=&{\varpi}^{\alpha \beta}
{\kappa}_{\beta}, \quad \quad {\varpi}^{\alpha \beta}=
{\omega}_n({\Gamma}^n)^{\alpha \beta}\\
{\delta}_{\kappa}x^n &=&-i{\delta}_{\kappa}\theta \Gamma^n \theta \\
{\delta}_{\kappa}e &=& 4ie\dot{\theta}\kappa
\eea
whre ${\kappa}$ is an anticommuting spinoral parameter. This symmetry allows
to reduce some of the $\theta$ - degrees of freedom (here half of them,
in general at most half) \cite{{sieg2},{sieg},{ev}}. \\
The first massive superparticle model which exhibits $\kappa$ - symmetry
was introduced in 1982 by de Azc\'{a}rraga and Lukierski \cite{al1}. In
their model this symmetry was firstly observed but the r\^{o}le of
such a gauge invariance in reduction of the degrees of freedom was first
pointed out by Siegel \cite{sieg} and he introduced modified action.
 To finally overcome the problem with
the fermionic kinetic term present in the Casalbuoni's model one has to
enlarge the superspace. In the de Azc\'{a}rraga-\-Lukierski model it is
done by considering the $N$-extended Minkowski superspace $(x_{\mu},
{\theta}_{i \alpha}, {\bar{\theta}}_{i}^{\dot{\alpha}})$, $i=1,2,...,N$
and introducing central charges to the superalgebra.
Hence the resulting underlying rigid symmetry gets enlarged to $N$-extended
super-Poincar\'{e} superalgebra. \\
The new "isotopic" structure allows to use internal symplectic metric
$A_{ij}=-A_{ji}$ and the expression of the form
${\theta}^{i \alpha}A_{ij}\frac{d}{d\tau}{\theta}^{j \alpha}$ now is
not a total time derivative and can contribute nontrivially to the
action. After obvious modification in the super one form
\beq
d{\omega}_{\mu}=dx_{\mu}-i(d{\theta}_i {\sigma}_{\mu} {\bar{\theta}}_i
-{\theta}_i \sigma_{\mu} d{\bar{\theta}}_i )
\eeq
the lagrangian function can be written as
\beq
L=-m\sqrt{{\dot{\omega}}_{\mu}{\dot{\omega}}^{\mu}} + i({\theta}^{i}_{\alpha}
A_{ij}{\dot{\theta}}^{j \alpha} + {\bar{\theta}}^{i}_{\dot{\alpha}}
A_{ij}{\dot{\bar{\theta}}}^{j \dot{\alpha}})
\eeq
The fermionic kinetic term is in fact of the Wess-Zumino type
and changes under supersymmetry transformations by a total time
derivative. Indeed, let $Z_{IJ}$ be a symmetric, Lorentz invariant
matrix (where $I$, $J$ could be multi-\-indices e.g.  $I=(\alpha , i)$
and $Z_{IJ}={\epsilon}_{\alpha \beta}A_{ij}$ ), then the simple example
of the WZ-term for a supersymmetric particle is of the form
\beq
S_{WZ}=\int d\tau i{\theta}^I Z_{IJ}{\theta}^J
\eeq
What means that one starts from the closed super-twoform $h=id\theta
Zd\theta$. It is exact; with $b=id\theta Z\theta$ one can write $h=db$.
From the invariance of of $h$ under supertranslations it follows that
$d({\delta}_{\epsilon}b)=0$ and at least locally
${\delta}_{\epsilon}b=df$ for some superfunction f. For the AL-action it
means that $\epsilon$-\-variation yields the total time derivative
change in the lagrangian.\\
The AL-model after the first quantization yields the irreducible
representations of the $N$-extended super-Poincar\'{e} superalgebra. The
whole spectrum of supersymmetric multiplets was found as a result of the
first quantization  \cite{{afl},{af1}} not only for the massive case but also for
the massless \cite{af2}. In the quantization of this model there was firstly
applied the supersymmetric generalization of the Gupta-\-Bleuler
\cite{{gup},{bleu}} quantization method \cite{{lus},{af1}}, which later was used
in quantization of various systems exhibiting the similar structure of
the second class constraints (i.e. hermitean splitting of the set of the
second class constraints into the subsets of conjugated, relatively
first class constraints).\\
The coupling of this model to the external
fields gives interesting results. Comparing to the traditional equations
of the spin precession in the external electromagnetic field we obtain
that the superspace generalization of the Bargmann-Michel-Telegdi equations
takes the form \cite{af3}
\bea
{\dot{p}}^{\mu}&=&eF^{\mu \nu} {\dot{x}}_{\nu} + g S_{\rho\lambda}^{Spin}
{\partial}^{\mu} F^{\rho\lambda} \\
\frac{d}{d\tau} W_{\mu}&=& gF_{\mu\nu}W^{\nu} +
(\frac{e}{2m}-g){\dot{\omega}}_{\nu}F^{\nu\rho}W_{\rho}{\dot{\omega}}_{\mu}-\\
&&- \frac{e}{2m}(W_{\nu}{\dot{\omega}}_{\mu}
-W_{\mu}{\dot{\omega}}_{\nu})F^{\nu\rho}{\dot{z}}_{\rho},
\eea
where $W_{\mu} = \frac{i}{2m^2}
{\varepsilon}_{\mu\nu\rho\lambda}p^{\nu}S_{Spin}^{\rho\lambda}$ and
$S^{Spin}_{\mu\nu}=
\frac{i}{2}{\varepsilon}_{\mu\nu\rho\lambda}p^{\rho}(i{\theta}_k
{\sigma}^{\lambda}{\bar{\theta}}_k )$. For the external Yang-Mills field
we obtain, within the minimal coupling, the generalized Wong equations
\cite{af3}
\bea
{\dot{p}}^{\mu}&=&g_1 F^{\mu \nu}_a I^a {\dot{x}}_{\nu} + g_2
S_{\rho\lambda}^{Spin}{\partial}^{\mu} F^{\rho\lambda}_a I^a \\
{\dot{I}}^a&=&g_1 f^a_{bc} A^b_{\mu}I^c {\dot{x}}^{\mu} - g_2 f^a_{bc}
F^b_{\mu\nu}I^c S^{\mu\nu}_{Spin}\\
\frac{d}{d\tau} W_{\mu}&=& g_2 F_{\mu\nu}^a W^{\nu}I_a +
(\frac{g_1}{2m}-g_2 ){\dot{\omega}}_{\nu}F^{\nu\rho}_a
W_{\rho}{\dot{\omega}}_{\mu}I^a  -\\
&&- \frac{g_1}{2m}(W_{\nu}{\dot{\omega}}_{\mu}
 -W_{\mu}{\dot{\omega}}_{\nu})F^{\nu\rho}_a {\dot{z}}_{\rho}I^a ,
\eea
However, the AL-model is supersymmetric therefore the fully
supersymmetric coupling to the supersymmetric field is of greater
interest. It can be found in Ref. \cite{lumil}. In the case of the
supersymmetric Yang-\-Mills and supergravity theories it gives in a
natural way the conventional sets of constraints for these fields.
\subsection{Twistorial models}
The supersymmetric particle models using the twistor-\-like variables
were developed in the eighties, firstly in the component formulation then in
the superfield one \cite{{bbcl},{pmc},{peri},{getk},{stv1},{stvz}}.
The very important result obtained within this formulation consists in
re-expressing upon use of the equations of motion the local world-line
supersymmetry as $\kappa$-transformation \cite{{stv1},{stvz}}.
General feature of this kind of models is a possibility of manifestly
covariant quantization. To merely signal the existence of very reach
developments let us recall the superfield version of the model beeing
generalization of the following component action \cite{stv1}
\beq
S= \int d\tau p_m ({\dot{x}}^m -i\bar{\theta}{\gamma}^m \theta
+\bar{\lambda} \gamma^m \lambda ).
\eeq
Namely,
\beq
S_1 = -i\int d\tau d\eta P_m (DX^m + i\bar{\Theta} {\gamma}^m D\Theta ),
\eeq
where $D={\partial}_{\eta} +i\eta {\partial}_{\tau}$ and $P_m = p_m
+i\eta {\rho}_m$, $X_m =x_m +i\eta {\chi}_m$, ${\Theta}_{\alpha} =
{\theta}_{alpha} +\eta {\lambda}_{\alpha}$. In the component version
this action contains additional to the $S$ an auxiliary term of the form
\beq
S_a = i\int d\tau {\rho}_m ({\chi}^m + \bar{\theta} {\gamma}^m \lambda )
\eeq
The mechanism of trading the twistorial superparticle's
$\kappa$-symmetry for world-line supersymmetry is analysed in series
of papers \cite{{stv1},{stvz},{bnsv}} and recently in Ref.
\cite{ghs}. Relation between the different forms of the superparticle
dynamics, involving spinorial coordinates is analysed in Ref. \cite{mc}.\\
The possibility of manifestly covariant quantization of the massless
particle model was the motivation of development of the model of
harmonic superparticle \cite{sok}. The action of this model is a
generalization of the Siegel model with some new (harmonic) bosonic
variables which are parametrising a suitably choosen coset spaces.
\subsection{Arbitrary superspin models}
The model of the classical arbitrary spin particle \cite{kuz}
discussed in Sec.2.2. can be generalized to the pseudoclassical model
with the N-\-extended super-Poincar\'{e} symmetry \cite{kuz2}. After the
Dirac quantization this model gives the on-shell massive chiral
superfields (the central charges can be introduced as well).\\
The extension of the configuration space $M\times S^2$ is done in the
Minkowski sector, it is changed into the N-extended superMinkowski
superspace with coordinates $( x^a , {\theta}^{\alpha I},
{\bar{\theta}}^{\dot{\alpha}}_I )$ $I=1,2,...,N$, $a=0,1,2,3$. On the
new configuration space $M^{4|4N}\times S^2$ there is defined the
Lagrangian of the form
\beq
L=\frac{1}{2}e^{-1}_1 ({\Pi}^2 - (e_1 m )^2 )+e^{-1}_2
(4\frac{z\dot{\bar{x}}}{(\Pi \cdot \xi )^2}e^2_1 + (\Delta e_2 )^2 ),
\eeq
where
\bea
{\Pi}^a &=& {\dot{x}}^a + i {\theta}^I {\sigma}^a{\dot{\bar{\theta}}}_I
- {\dot{\theta}}^I {\sigma}^a {\theta}_I \\
{\xi}_a &=& ({\sigma}_a )_{{\alpha} \dot{\beta}}
z^{\alpha}{\bar{z}}^{\dot{\beta}} \\
\Delta &=& m\sqrt{Y(Y+1)}
\eea
The $Y$ is a superspin parameter. The central charges analogous to those
of the Azc\'{a}rraga-\-Lukierski model can be considered here as well
\cite{kuz2}.
\section{Doubly supersymmetric models}
The doubly supersymmetric models were considered firstly by Gates and
Nishino \cite{gat1}. To extend the NSR string theory they proposed a new
class of superstring models which possess both spacetime and world-sheet
supersymmetries. Then within this scheme the particle model was
considered \cite{{gat2},{gat3}}. There are also two other approaches to
such particle models: the twistor-like superfield models (commented in
previous section) and the spinning particle model (invented firstly in
the component form \cite{{aklh},{kgl}}).\\
The spinning particle models are revieved in Ref. \cite{flr}, here we
shall restrict ourselves to the brief ilustration of the superfield
realization \cite{{kam},{af4}}.
In the (supersymmetry)$^2$ particle models one introduces the supertime
space $(t,\eta )$ and the superMinkowski superspace. Therefore the
trajectories of a point object are the mappings
\beq
(t ,\eta ) \in T \mapsto M_D \ni (X^m , {\Theta}^A )
\eeq
where
\bea
X^m (t,\eta ) &=& x^m (t) +\eta{\Lambda}^m (t)\\
{\Theta}^A (t,\eta ) &=& {\theta}^A (t) + \eta {\varphi}^A
\eea
Introducing the covariant object
\beq
Y^m = {\nabla}_{\eta} X^m + i\Theta {\gamma}^m {\nabla}_{\eta} \Theta ,
\eeq
where $\nabla$ is the covariant superderivative given by ${\nabla}_A
=E_A^M {\partial}_M$ (cf. eq. (71)) one can write the supersymmetric
invariant action in the form
\bea
S&=&\frac{1}{2}\int dt d\eta sdet(E_A^M ){\nabla}_{\eta} Y^m \cdot Y_m
=\\
&=& \frac{1}{2}\int dt ( e^{-1} {\dot{\omega}}^2 - 2e^{-1} \psi
{\lambda}^m {\dot{\omega}}_m -2i({\dot{\omega}}^m -\psi {\lambda}^m
)\bar{\varphi}{\gamma}_m \varphi -\\
&&-e( \varphi {\gamma}^m \varphi )^2+\lambda \dot{\lambda} )
\eea
The superfield covariant phase space description of this model in the
rigid supersymmetry case was given in Ref. \cite{afl}.\\
One can say that developments in the spinning
and superparticle models has been resumed
in their superfield formulation which appeared in the second half of the
eighties. It has turned out that all types of the pseudomechanical
description can be put together and organized in a joint superfield
model (let us note that there exists the superfield formulation of a
spinning particle alone, but not of the superparticle, which has to coexist
in the superfield formulation with the spinning particle).
\section{Recent developments: q-deformed spinning particle
and $\kappa$-relativistic model}
Finally let us mention the brand new aspect of the relativistic particle
models, namely their deformations. Actually there are not well
established deformed models. However, without entering into the question
why things have to be (or not to be) deformed we shall recall two examples:
the q-deformed spinning particle and the $\kappa$-relativistic particle.\\
The example of the q-deformed relativistic spinning particle was considered
by Malik \cite{mal}. With the use of the first order Lagrangian of the
spinning particle and the q-deformed  graded commutation relations for $(x^m
,{\psi}^m ,p^m ,e)$ in the phase space he introduces "deformed"
GL$_q$(2)-invariant Lagrangian
\beq
L=\sqrt{q} p_m{\dot{x}}^m + \frac{i}{2}{\psi}^m{\dot{\psi}}_m
-\frac{e}{1+q^2}p^2 +i\chi{\psi}_m p^m
\eeq
The model is under investigation and its Dirac "deformed" quantization
is still to be performed.\\
The $\kappa$-relativistic particle is more "physical". It lives in the
$\kappa$-deformed Minkowski space \cite{{lnr1},{lnr2}} with the mass
shell condition modified to the following form
\beq
(2\kappa \sinh \frac{p_0}{2\kappa} )^2 -{\vec{p}}^2 = m^2
\eeq
This model can be described within the formalism with commuting as well
as noncommuting space-time coordinates. The interesting properties of
the object of this kind are discussed in Refs. \cite{{lrz},{lrz1}}.
\section*{Acknowledgements}
This work is supported in part by KBN Grant \# 2 P302 087 06.

\vspace*{1cm}
{\footnotesize January 1996, ITP UWr 901/96}

\begin{thebibliography}{99}
\bibitem{fl} Frenkel J., Z. f\"{u}r Physik {\bf 37} (1926), 243
\bibitem{tho} Thomas L.H., Phil.Magazine {\bf 3} (1927), 1
\bibitem{bmt} Bargmann V., Michel L., Telegdi V.L., Phys. Rev. Lett.
{\bf 2} (1959), 435
\bibitem{bar} Barut A.O., "{\em Electrodynamics and Classical Theory of
Fields and Particles"}, MacMillan, New York 1964
\bibitem{hr} Hanson A.J., Regge T., Ann.Phys.  {\bf 87} (1974), 498
\bibitem{hrt}  Hanson A.J., Regge T., Teitelboim C., {\em Constrained
Hamiltonian Systems"}, Accad.Naz.dei Lincei, Rome 1976
\bibitem{gra} Grassberger P., J.Phys. A: Math Gen. {\bf 11} (1978), 1221
\bibitem{csvz} Cognola G., Soldati R., Vanzo L., Zerbini S., Phys. Lett.
{\bf 104 B} (1981), 67
\bibitem{mbss} Balachandran A.P., Marmo G., Stern A., Skagerstam Bo-S.,
Phys. Lett. {\bf 89 B} (1980), 199
\bibitem{ho1} Cho J-H., Hyun S., Kim J-K., {\em "A Covariant Formulation
of Classical Spinning Particle"}, preprint YUMS-93-09, Seoul 1993
\bibitem{csz}  Cognola G., Soldati R., Zerbini S., preprint
UTF77, Univ. di Trento, 1982
\bibitem{ss} Stern A., Skagerstam Bo-S., Physica Scripta {\bf 24}
(1981), 493
\bibitem{cas} Casalbuoni R., Nuovo Cim. {\bf 33 A} (1976), 389
\bibitem{alm} Almond P., {\em "The Suprsymmetry Extended Weyl Algebra
and Casalbuoni's G$_4$ Model"}, preprint QMC/02-81, London 1981
\bibitem{bap} Barut A.O.,Pav\u{s}i\u{c} M., Phys. Lett. {\bf B
216} (1989), 297
\bibitem{ho2} Cho J-H., Hyun S., Kim J-K., {\em "Relation between
Classical and Pseudo-classical Spinnig Particle"}, preprint YUMS-93-8,
Seoul 1993
\bibitem{bm}  Berezin F.A., Marinov M.S., Ann. Phys. {\bf 104} (1977),
336
\bibitem{bcl} Barducci A., Casalbuoni R., Lusanna L., Nuovo Cim. {\bf 35
A} (1976), 377
\bibitem{ma} Martin J., Proc. Roy. Soc. of London {\bf A 251} (1959),
536
\bibitem{ms} Matthews P., Salam A., Nuovo Cim.  {\bf 2} (1955), 120
\bibitem{tob} Tobocman W., Nuovo Cim. {\bf 3} (1956), 134
\bibitem{vp} Volkov D.V., Pashnev A.T., Teor. Mat. Fiz. {\bf 44}
(1980), 321
\bibitem{al1} de Azc\'{a}rraga J.A., Lukierski J., Phys. Lett. {\bf 113
B} (1982), 170
\bibitem{dix} Dixon W.G., Nuovo. Cim. {\bf 34} (1964), 317
\bibitem{bdh} Brink L., Di Vecchia P., Howe P., Nucl. Phys. {\bf B 118}
(1977), 76
\bibitem{bdz} Brink L., Deser S., Zumino B., Di Vecchia P., Howe P.,
Phys. Lett. {\bf 64 B} (1976), 437
\bibitem{bs} Brink L., Schwarz J.H., Phys. Lett. {\bf 100 B} (1981),
310
\bibitem{gt} Galvao C.A.P., Teitelboim C., J. Math. Phys. {\bf 21}
(1980), 1863
\bibitem{sund} Sundermeyer K., {\em "Constrained Dynamics"} Springer,
Berlin 1982
\bibitem{tow} Townsend P.K.,{\em "Spacetime supersymmetric particles and
strings in background fields"} in proceedings of the first Torino meeting
on Superunification and Extra Dimensions,1985
\bibitem{sieg1} Siegel W., Class. Quantum Grav. {\bf 2} (1985). L95
\bibitem{sieg2} Siegel W., Phys. Lett. {\bf 203 B} (1988), 79
\bibitem{sieg} Siegel W., {\em "Introduction to string field theory"},
 World Scientific, Singapore 1988
\bibitem{ev} Evans J.M., Nucl. Phys. {\bf B 331} (1990), 711
\bibitem{bggs} Barducci A., Giachetti R., Gomis J., Sorace E., J. Phys
A: Math. Gen. {\bf 17} (1984), 3277
\bibitem{gup} Gupta S., Proc. Roy. Soc. of London {\bf 63 A} (1950), 681
\bibitem{bleu} Bleuler K., Helv. Phys. Acta. {\bf 23} (1950), 567
\bibitem{lus} Lusanna L., in {\em "Supersymmetry and Supergravity"}
proceedings of XIX Karpacz Winter School of
Theor. Phys. World Scientific, Singapore 1983
\bibitem{af1} Frydryszak A., Phys. Rev. {\bf D 30} (1984), 2172
\bibitem{af2} Frydryszak A., Phys. Rev. {\bf D 35} (1987), 2432
\bibitem{af3} Frydryszak A.,{\em "Supersymmetric particle with internal
symmetries in external electromagnetic and Yang-\-Mills fields"}
IFT/UWr 1982
\bibitem{al2} de Azc\'{a}rraga J.A., Lukierski J., Phys. Rev. {\bf D 28}
(1982), 1337
\bibitem{afl}  Frydryszak A., Lukierski J., Phys. Lett. {\bf 117B}
(1982), 51
\bibitem{1001} Gates Jr. S.J., Grisaru M.T., Rocek M., Gates W., {\em
"Superspace"} Benjamin/Cummings, London 1983
\bibitem{ike1} Ikemori H., {\em "Superfield formulation of
superparticle"} \# DPNU-88-03, Nagoya Univ. preprint, 1988
\bibitem{ike2} Ikemori H., Z. f\"{u}r Physik {\bf C}: Particles and
Fields {\bf 44} (1989), 625
\bibitem{gat1} Gates S.J. Jr., Nishino H., Class. Quantum. Grav. {\bf 3}
(1986), 745
\bibitem{gat2} Gates S.J. Jr., Majumdar P., Mod. Phys. Lett. {\bf 4A}
(1989), 339
\bibitem{gat3} Gates S.J. Jr., in {\em "Functional integration, geometry
and strings"}, proceedings of XXV Karpacz Winter School of Theor. Phys.,
Birkh\"{a}user, Basel 1989
\bibitem{gatn1} Gates S.J. Jr., Rana L., {\em "A Theory of Spinning Particles
for Large N-extended Supersymmmetry "} hep-th/9504025
\bibitem{gatn2} Gates S.J. Jr., Rana , {\em "A Theory of Spinning Particles
for Large N-extended Supersymmmetry (II)"} hep-th/9510151
\bibitem{sou} Souriau J.-M., {\em "Structure des syst\`{e}mes
dynamiques"}, Dunod, Paris 1970
\bibitem{zakrz} Zakrzewski S., {\em "Extended phase space for a spinning
particle"}, hep-th/9412100
\bibitem{duvhor} Duval Ch., Horvathy P., Ann. Phys. (NY) {\bf 142}
(1982), 10
\bibitem{duv1} Duval Ch., Ann. Inst. H. Poincar\'{e}  {\bf A} XXV
(1976), 345
\bibitem{lumil} Lusanna L., Milewski B., Nucl. Phys. {\bf B 247} (1984),
396
\bibitem{bbcl} Bengtsson A.K.H., Bengtsson I., Cederwall M., Linden N.,
Phys. Rev. {\bf D 36} (1987), 1766
\bibitem{pmc} Penrose R., Mac Callum M.A.H., Phys. Rep. {\bf 6} (1973),
109
\bibitem{has} Hasiewicz Z., Siemion P., Defever F.,
Int. J. Mod. Phys. A {\bf 17} (1992), 3979
\bibitem{kuz} Kuzenko S.M., Lyakhovich S.L., Segal A.Yu., Int. J. Mod.
Phys. {\bf A 10} (1995), 1529
\bibitem{peri} Penrose R., Rindler W., {\em "Spinors and space-time"},
Cambridge Univ. Press, Cambridge 1986
\bibitem{getk} Gershun V.D., Tkach V.I., JETP Lett. {\bf 29} (1979), 320
\bibitem{hppt} Howe P.S., Penati S., Pernici M., Townsend P., Phys.
Lett. {\bf B 215} (1988), 255
\bibitem{fer} Ferber A., Nucl. Phys. {\bf B 132} (1977), 55
\bibitem{shi} Shirafuji T., Prog. Theor. Phys. {\bf 70} (1983), 18
\bibitem{kuz2} Kuzenko S.M., Lyakhovich S.L., Segal A.Yu., Phys. Lett.
{\bf B 348} (1995), 421
\bibitem{stv1} Sorokin D.P., Tkach V.I., Volkov D.V., Mod. Phys. Lett.
{\bf A4} (1989), 901
\bibitem{stvz} Sorokin D.P., Tkach V.I., Volkov D.V., Zheltukhin A.A.,
Phys. Lett. 216{\bf B} (1989), 302
\bibitem{bnsv} Bandos I.A., Nurmagambetov A., Sorokin D.P., Volkov D.V.,
Class. Quantum. Grav. {\bf 12} (1995), 1881
\bibitem{ghs} Galperin A.S., Howe P.S., Stelle K.S., Nucl. Phys. {\bf B
368} (1992), 248
\bibitem{mc} Cederwall M., {\em "A note on the Relation between
Different Forms of Superparticle Dynamics"}, preprint ITP-93-33,
G\"{o}teborg 1993 (hep-th/9310177)
\bibitem{sok} Sokatchev E., Class. Quantum. Grav. {\bf 4} (1987), 237
\bibitem{aklh} Aoyama S., Kowalski-Glikman J., Lukierski J., van Holten
J.W., Phys. Lett. {\bf 217B} (1989), 95
\bibitem{kgl} Kowalski-Glikman J., Lukierski J., Mod. Phys. Lett. {\bf
A4} (1989), 2437
\bibitem{kam} Kavalov A., Mkrtchyan R.L., {\em "Spinning
superparticle"}, preprint Yer PhI/1068(31)-88, Yerevan, 1988
(unpublished)
\bibitem{af4} Frydryszak A., {\em "Superfield Spinning Superparticle
Model}\\{\em and (Supersymmetry)}$^2${\em "}, preprint ITP UWr 726/89,
Wroc{\l}aw, 1989 (unpublished)
\bibitem{afl} de Azc\'{a}rraga J.A., Frydryszak A., Lukierski J., Phys.
Lett. {\bf B 247} (1990), 289
\bibitem{flr} Frydryszak A., Lukierski J., {\em "Spinning Superparticle
Models - Recent Developments"}, preprint  ITP/UWr 752/90, Wroc{\l}aw
1990
\bibitem{mal} Malik R.P., {\em "On q-deformed spinning relativistic
particle"}, hep-th/950302
\bibitem{lrz} Lukierski J., Ruegg H., Zakrzewski W.J., in {\em "Quantum
Groups. Formalism and Applications"}
proceedings of XXX Karpacz School of Theor. Phys., PWN, Wroc{\l}aw 1995
\bibitem{lrz1} Lukierski J., Ruegg H., Zakrzewski W.J., Ann. Phys. {\bf
243} (1995), 90
\bibitem{lnr1} Lukierski J., Nowicki A., Ruegg H. Phys. Lett. {\bf B
264} (1991), 331
\bibitem{lnr2} Lukierski J., Nowicki A., Ruegg H. Phys. Lett.  {\bf B
313} (1993), 357
\end{thebibliography}
\end{document}